# Dynamical Control of Excitons in Atomically Thin Semiconductors

Eric L. Peterson[1*], Trond I. Andersen[1,2*], Giovanni Scuri[1,3], Andrew Y. Joe[1,4], Andrés M. Mier Valdivia[5], Xiaoling Liu[1], Alexander A. Zibrov[1], Bumho Kim[6], Takashi Taniguchi[7], Kenji Watanabe[8], James Hone[9], Valentin Walther[1,10,11], Hongkun Park[1,12], Philip Kim[1,5], Mikhail D. Lukin[1,†]

[1]Department of Physics, Harvard University, Cambridge, MA 02138, USA
[2]Google Research, Mountain View, CA 94043, USA
[3]E. L. Ginzton Laboratory, Stanford University, Stanford, CA 94305, USA
[4]Department of Physics and Astronomy, University of California, Riverside, CA 92521, USA
[5]John A. Paulson School of Engineering and Applied Sciences, Harvard University, Cambridge, MA 02138, USA
[6]Department of Physics and Astronomy, University of Pennsylvania, Philadelphia, PA 19104, USA
[7]International Center for Materials Nanoarchitectonics, National Institute for Materials Science, 1-1 Namiki, Tsukuba 305-0044, Japan
[8]Research Center for Functional Materials, National Institute for Materials Science, 1-1 Namiki, Tsukuba 305-0044, Japan
[9]Department of Mechanical Engineering, Columbia University, New York, NY 10027, USA
[10]Department of Chemistry, Purdue University, West Lafayette, IN 47906, US
[11]Department of Physics and Astronomy, Purdue University, West Lafayette, IN 47906, USA
[12]Department of Chemistry and Chemical Biology, Harvard University, Cambridge, MA 02138, USA

*These authors contributed equally to this work.
†To whom correspondence should be addressed: lukin@physics.harvard.edu

**Excitons in transition metal dichalcogenides (TMDs) have emerged as a promising platform for novel applications ranging from optoelectronic devices to quantum optics and solid state quantum simulators. While much progress has been made towards characterizing and controlling excitons in TMDs, manipulating their properties during the course of their lifetime — a key requirement for many optoelectronic device and information processing modalities — remains an outstanding challenge. Here we combine long-lived interlayer excitons in angle-aligned MoSe$_2$/WSe$_2$ heterostructures with fast electrical control to realize dynamical control schemes, in which exciton properties are not predetermined at the time of**

**excitation but can be dynamically manipulated during their lifetime. Leveraging the out-of-plane exciton dipole moment, we use electric fields to demonstrate dynamical control over the exciton emission wavelength. Moreover, employing a patterned gate geometry, we demonstrate rapid local sample doping and toggling of the radiative decay rate through exciton-charge interactions during the exciton lifetime. Spatially mapping the exciton response reveals charge redistribution, offering a novel probe of electronic transport in twisted TMD heterostructures. Our results establish the feasibility of dynamical exciton control schemes, unlocking new directions for exciton-based information processing and optoelectronic devices, and the realization of excitonic phenomena in TMDs.**

Techniques for controlling excitons in two-dimensional semiconductor systems have been actively explored for the past three decades[1,2]. In the pioneering work involving semiconductor double quantum wells (DQWs), the ability to control spatially indirect excitons through electrostatic gating enabled the realization of elementary excitonic circuits[3] and the production of trapped, spatially coherent exciton states[4]. More recently, similar techniques have been realized with excitons in van der Waals TMD heterostructures[5,6]. The unique properties of excitons in TMDs have extended the range of possibilities for exciton gate tunability, including demonstrations of electrically-switchable two-dimensional mirror and beam-steering devices[7–10] as well as valleytronic switches[11,12]. Moreover, combining electrical control with the ability to layer TMDs in different material combinations and at controlled relative twist angles has led to the realization of new excitonic phenomena such as layer-hybridized excitons[13,14], gate-tunable moiré excitons[15,16], and strongly correlated excitonic insulators[17–19].

While the electrical control of excitons in TMDs has so far been restricted to their steady-state response or dynamics as predetermined at the time of optical excitation, the ability to leverage gate control during the exciton lifetime[3,20,21] can open up fundamentally new opportunities and applications. Specifically, a number of modern applications in quantum optics, quantum simulation and quantum information science utilize dynamical control schemes, in which the properties of emitters or excitations are manipulated after creation. For example, the ability to execute a programmable control sequence within the emitter lifetime (as illustrated in Figure 1a) is essential for the storage and retrieval of light pulses[22–24], and the generation of complex non-classical states of light and matter[25–27]. In the context of quantum simulation, trapping potentials and other control parameters are commonly varied in order to prepare quantum phases of matter[28–30] and study their non-equilibrium dynamics[31,32]. Additionally, the ability to manipulate excitons within their lifetime is a necessary ingredient in performing the sequences of conditional operations that comprise logical circuits and programs, which form the foundation of classical and quantum information processing. While this type of control could unlock a vast range of possibilities if realized in TMDs, such capabilities have so far been limited to more traditional exciton systems[3,20,21] as TMDs typically feature very short exciton lifetimes (picoseconds in monolayers[33] and femtoseconds in natural bilayers[34–36]) and a slow rate of gate operation due to the formation of large-resistance Schottky barriers at the metal-TMD interface, with seconds-scale charge injection timescales having been reported in the literature[37,38].

Here we address this challenge by combining fast electrical control with long-lived interlayer excitons (IEs) in a locally-gated field-effect transistor (FET) device. Specifically, we employ a device consisting of $MoSe_2/WSe_2$ heterobilayers stacked with a marginal (less than 2°) twist angle and encapsulated in hBN, with patterned metallic top and bottom gates (Fig. 1b). Due to the type-II band alignment of $MoSe_2$ and $WSe_2$, optical excitation of the heterobilayer produces interlayer excitons, in which the electrons and holes are localized in the $MoSe_2$ and $WSe_2$ layers respectively. The formation of IEs occurs on ultrafast (femtosecond) timescales[39], and due to the small wavefunction overlap between the electron and hole relative to intralayer excitons, such excitons can possess lifetimes approaching microseconds[40–42]. Importantly, interlayer excitons feature an out-of-plane electric dipole moment that allows their lifetime and emission energy to be tuned through an out-of-plane electric field on timescales determined by charge injection into the metallic gates rather than the semiconducting TMDs. The patterned gate geometry of the device additionally allows charge to be transferred between the gated and ungated regions on timescales determined by the resistance and capacitance of the heterobilayer system itself, rather than the electrical contacts[37]. This combination enables both the electric field and the local doping to be modulated during the exciton lifetime, in stark contrast to the control schemes employed in TMDs thus far, where the exciton response is pre-determined at the time of excitation.

**Long-lived interlayer excitons in $MoSe_2$/$WSe_2$ heterobilayers**

We first characterize the excitons and their gate tunability in our device. Figure 1c shows the interlayer exciton PL spectrum with both gates grounded. We attribute the two prominent peaks, split by about 20 meV, to emission from singlet and triplet interlayer excitons[12] which correspond to transitions from the upper and lower $MoSe_2$ conduction bands (Fig. 1c, inset). A less prominent low-energy peak, which could be a phonon replica[43,44] or emission from a moiré-trapped state[45], is also visible. The emission energies are widely tunable using vertical electric fields, produced by applying opposite voltages to the top and bottom gates of the heterostructure (Fig.

1d). Comparing the peak energy shift to the linear Stark formula, $-edE_{hs}$ ($e$ is the electron charge and $d$ is the electron-hole separation; $E_{hs}$ is the electric field across the TMD heterostructure), we extract an IE electric dipole moment $p = ed \approx e \times 0.52$ nm, which is similar to the typical separation (0.6 nm) of the MoSe$_2$ and WSe$_2$ layers. The exciton properties in our sample can additionally be tuned by applying a positive voltage to only one of the top or bottom gates, which electrostatically n-dopes the MoSe$_2$ layer (Fig. 1e) and promotes the formation of charged interlayer excitons associated with red-shifted emission[12,40].

Time-resolved PL measurements reveal that the IE lifetime is many orders of magnitude longer than typical intralayer exciton lifetimes, with radiative dynamics that are gate-tunable (Fig. 1f). We fit exciton lifetime measurements with the top gate voltage held at $V_T = \pm 2$ V (corresponding to charged and neutral IE states) to a phenomenological biexponential decay model and extract a slow decay timescale of 219 ns for $V_T = +2$ V and 367 ns for $V_T = -2$ V (the fast timescale $\tau_1 \approx 26$ ns does not vary strongly with gate voltage; see Supplementary Note I and Supplementary Fig. S1 for further discussion of the PL decay curves). Notably, we find that the charged IE possesses both a shorter total lifetime and reduced emission intensity compared to the neutral IE. This can be understood by considering both the radiative and non-radiative effects governing exciton lifetime: First, doping the sample reduces the oscillator strength of the radiative transition[46], thus increasing the radiative lifetime $\tau_{rad}$ and reducing the PL intensity. In addition, doping is expected to open additional non-radiative decay channels through scattering with free carriers[47] and shorten the non-radiative lifetime $\tau_{nr}$. Considering both effects in combination and noting that the total lifetime of the charged IE is shorter despite its lower radiative rate, we conclude that the slow timescale $\tau_2$ is likely dominated by non-radiative processes.

**Dynamical exciton control**

Having established the tunability of the exciton energy and radiative dynamics in our device, we turn to the realization of dynamical exciton control -- specifically, modulation of the interlayer exciton emission energy during its lifetime. We perform time-resolved PL measurements while exciting the sample with a pulsed laser, which is synchronized to an arbitrary waveform generator (AWG) connected to the device gates (Fig. 1g). This allows us to apply programmable waveforms and pulse sequences to the device at well-defined times after excitation. To modulate the PL energy, we use the AWG to apply a vertical electric field whose direction is reversed during the exciton lifetime, tuning the energy through the linear Stark shift. The pulse sequence for this measurement is shown in Fig. 2a. We measure the change in the PL emission spectrum throughout the IE lifetime by sweeping a tunable filter across the resonance and performing a time-resolved PL intensity measurement at each filter position (Supplementary Note II and Supplementary Fig. S2). Indeed, a plot of the exciton spectrum as a function of time (Fig. 2b) shows a large change in peak emission wavelength subject to the dynamical electric field reversal. Fitting the PL spectra integrated before and after the electric field direction is reversed (Fig. 2c), we extract an energy modulation of ~9 meV, in excellent agreement with the static gate-dependence. We note that the magnitude of the Stark shift realized in this measurement is limited by the output voltage range of our AWG – gate-induced shifts exceeding 200 meV have been demonstrated in DC experiments[40].

The exciton response to a step function voltage also reveals the timescale over which our device responds to fast control signals. In particular, the step function voltage evidently produces a damped oscillation ("ringdown") in the PL energy (Fig. 2b), which we attribute to resonances in the high-frequency transmission lines leading to the top gate of the device. The ringdown frequency is about 80 MHz, with a damping time of about 15 ns (see Supplemental Notes III and IV and Supplementary Figs. S3-4 for decay curve fits and characterization of high-frequency electrical transmission). The short transient response time confirms that electrostatic gating can be used to modulate interlayer exciton properties on timescales more than an order of magnitude faster than their lifetime.

The time-resolved spectra also reveal blueshifted emission at times shortly after excitation, which becomes more pronounced with increased laser power (Fig. 2d). This feature is similar to that previously observed in GaAs double quantum wells[48] and TMD heterobilayers[40]. This nonlinear blueshift arises from density-dependent repulsive dipolar interactions, but in the steady state is typically only appreciable for large optical excitation powers. In our experiment, the strength of the nonlinear blueshift is greater than the IE linewidth even for relatively low pulsed excitation powers (~1 μW, with a repetition rate of 2.11 MHz and pulse width 10 ps) when the exciton density is largest. This regime could be interesting for future study, as the large blueshift at low to moderate excitation powers is reminiscent of the strong nonlinearity associated e.g. with the dipole blockade phenomenon[49–51].

**Electron density modulation**

Next, we turn to the case where only one gate is modulated. While the rapid response of the exciton PL energy to changes in electric field can be understood as an extension of the steady-state gate tunability, modulating a single gate to tune the electron density leads to more dramatic and surprising behavior. Toggling the top gate between $V_T = +2$ (electron-doped) and $V_T = -2$ (intrinsic) for 50 ns roughly 150 ns after the excitons are created, we observe a large response in the PL intensity that becomes highly pronounced at low optical excitation power (Fig. 3a). Decreasing the laser excitation power from a few microwatts to tens of nanowatts has two effects: it reduces the prominence of a fast initial decay component (Fig. 3a, left inset) which is attributed to interaction-driven diffusion[40], and it evidently enhances the PL intensity contrast between the two voltages from a factor of ~2 to a factor ~7 (Fig. 3a, right inset). The response at low excitation powers is so substantial that it becomes possible to toggle the PL intensity to levels above the initial emission rate, even hundreds of nanoseconds after excitation (Fig. 3b). Despite the inefficiency of charge injection through the TMD contacts, this behavior can be achieved with gate pulses as short as 10 ns (see Supplemental Note V and Supplemental Fig. S5 for demonstration of similar control timescales in a second device). We observe that for this sample location and pulse sequence, the gate modulation affects the exciton radiative rate much more strongly than its total lifetime, as indicated by the similar PL intensities after 50 ns and 200 ns gate pulses (orange and blue curves in Fig. 3a). Other sample locations showed reduced PL intensity after $V_T$ was returned to +2 V; the variations in response are likely caused in part by twist angle inhomogeneity, an effect that could be alleviated in heterobilayers with larger lattice mismatch, such as $WSe_2/WS_2$[42,52,53].

**Spatial dependence of exciton response to gate modulation**

To better understand the origin of these observations, we studied the spatial dependence of the modulation response across the local gate arrays in the device (Fig. 4a). By exciting the sample globally with a defocused optical pulse and modulating the top gates with a sinusoidal waveform (here, 80 MHz) while the bottom gate is kept grounded, we measure the exciton response as quantified by the real part of the Fourier transform at the gate drive frequency, i.e. Re[$F(\mathbf{r}, \omega_0 = 2\pi \cdot 80$ MHz)] (Figs. 4b, c). In fact, we find that excitons at locations outside of the gates respond 180° "out-of-phase" with those underneath the gates.

We interpret the primary mechanism behind our observations to be toggling of the exciton charge state, with charge being pulled from outside the narrow patterned gates of the device (Fig. 4d). The fact that we see stronger out-of-phase response from some areas than others is likely due to the presence of charge traps in the heterostructure. It has been noted in the context of quasistatic optically-detected resistance and capacitance measurements[37] that for gate frequencies above tens of Hz, TMD layers essentially become electrically floated due to Schottky barriers at the contact-TMD junctions. As a result, the effect of the time-varying gate voltage is to redistribute charge across the device. This phenomenon has been used to circumvent poor electrical contacts in studies of electronic phenomena in TMDs, but has not yet been extended to MHz timescales or exploited for exciton control.

The above explanation captures the key aspects of the spatiotemporal phenomena in Figures 3 and 4. Notably, we can exclude several other potential mechanisms by comparing their expected effects with our data. While an electric field $E_{hs}$ can in principle affect the PL intensity through its effect on the radiative decay rate $\gamma_{\text{rad}}$ ($d\gamma_{\text{rad}}/dE_{hs} < 0$), considering that the sign of the exciton response does not change when applying the pulse to the bottom gate rather than the top gate, we can rule out electric field tunability as the primary mechanism underlying the response (Fig. 4e). While it could also be natural to consider the possibility of exciton motion induced by a spatially-varying potential due to the Stark shift $U = -p \cdot E_{hs}$ and/or Coulomb attraction to the gates, both of these scenarios would lead to PL responses that are opposite to those we observe in the experiment (Fig. 4f). Naturally, the ungated regions show no corresponding static response to changes in the gate voltages, implying that their opposite dynamical response is unique to the case where the gate is modulated quickly (Supplementary Note VI and Supplementary Fig. S6). At the same time, the redistribution of charge between gated and ungated regions (Fig. 4g) is consistent with our data and explains the opposite and apparently correlated responses between such neighboring regions, as well as the existence of a nanosecond-scale voltage response in regions that show no DC response.

**Conclusions**

Our experiments demonstrate dynamical control of interlayer exciton emission wavelength and intensity by modulating their interactions with applied electric fields and charge carriers on timescales much faster than their radiative dynamics. The ability to control excitons during the course of their lifetime constitutes a new modality for TMD-based optical devices and opens many new directions for realizing excitonic phenomena and device applications in TMDs. For example, tuning excitons into resonance with other emitters, or with photonic structures, could lead to novel photonic quantum systems with controlled interaction dynamics. Additionally, modulation of long-lived excitons could extend exciton-based sensing techniques to light-sensitive phenomena, such as fragile correlated electronic states in moiré semiconductor superlattices, by enabling one to study the exciton response long after optical excitation has ceased. Moreover, the valley selection rules of excitons in $MoSe_2/WSe_2$ have been shown to be "switchable" with doping[11,12], hence toggling the exciton charge state within the valley lifetime of ~40 ns[54] is an important step towards valleytronic information processing. Finally, tuning the radiative rate of long-lived excitons during their lifetime could be viewed as a step towards realizing photonic storage and retrieval schemes[23,24], in which incoming photons are stored in the device as excitons and controllably released on demand at a later time through application of a gate voltage. The photoluminescence experiments presented here are based on off-resonant pumping of excitons and collection of their subsequent spontaneous emission, which is incoherent and therefore cannot itself be a means of photonic storage and retrieval. However, new methods to resonantly couple light to long-lived excitons could potentially be explored for example, by utilizing layer-hybridized excitons[13,14,49,55], or by miniaturizing the patterned gates to subwavelength scales in order to couple to subradiant states[25,56,57]. By combining such methods with the technique demonstrated here, one could enable realization of on-chip, voltage-controlled optical delay lines at the ultimate miniaturization limit, or new approaches to quantum networks.

## Methods
### Sample preparation

The device studied in the main text has been described in previous work[12]. The thin hBN layers and monolayers of $MoSe_2$ and $WSe_2$ were mechanically exfoliated from bulk crystals onto Si substrates coated with 285 nm-thick $SiO_2$. The thicknesses of the hBN layers (100 nm and 70 nm for the top and bottom layers, respectively) were measured using atomic force microscopy. The monolayer TMDs were identified using an optical microscope and verified with photoluminescence measurements.

The device was assembled beginning with patterned Cr (1 nm) / PdAu (9 nm) bottom gates deposited on a $Si/SiO_2$ (285 nm) wafer. The bottom hBN layer was then transferred onto the bottom gates using the dry transfer method[58]. Pre-patterned Cr / Pt contacts were then deposited onto the bottom hBN layer. The top hBN and TMD layers were subsequently assembled using the same dry transfer technique and transferred onto the Pt contacts. The TMD layers were aligned by their natural edges, with a relative twist angle of $\Delta\theta \approx 58.8°$. Finally, the Cr (1 nm) / PdAu (9 nm) top and contact gates were deposited along with Cr (5 nm) / Au (120 nm) bonding pads and connections to the gates and contacts.

The $MoSe_2$ contact gates are placed on the side, some distance away from the excitation spot. Instead, the patterned top gates provide sufficient electron doping in the contact region. The $WSe_2$ contact gates were defective in these measurements, which is likely the reason for inefficient hole doping.

Experimental methods

All measurements were conducted in a 4 K cryostat (Montana Instruments) using a self-built confocal setup with a Zeiss (100x, NA = 0.75, WD = 4 mm) objective. The optical excitation and collection spots were scanned independently using two pairs of galvo mirrors. Photoluminescence spectrum measurements were carried out using a pulsed supercontinuum laser from NKT Photonics and a spectrometer. For the time-resolved PL intensity measurements, a Time Correlated Single-Photon Counting system (PicoHarp 300) was used to time-bin collected photons from an avalanche photodetector (APD), and the sample was excited using the supercontinuum laser operating at its minimum pulse repetition rate of 2.11 MHz. Interlayer exciton photoluminescence was spectrally filtered using tunable shortpass and longpass filters (Semrock TLP01-887 and TSP01-995). DC voltages were applied to the gates using Keithley 2400 multimeters, while AC voltages were applied using an arbitrary waveform generator (Tektronix AWG710).

**Acknowledgments**
We thank A. Sushko and P. Dolgirev for helpful discussions. We acknowledge support from the DoD Vannevar Bush Faculty Fellowship (N00014-16-1-2825 for H.P., N00014-18-1-2877 for P.K.), NSF (PHY-2012023 for H.P. and M.D.L.), Center for Ultracold Atoms (an NSF Physics Frontier Center) (PHY-1734011 for H.P. and M.D.L.), AFOSR (FA2386-21-1-4086 for P.K.), the Partnership for Quantum Networking from Amazon Web Services (A50791 for H.P. and M.D.L), and Samsung Electronics (for P.K. and H.P.). All fabrication was performed at the Center for Nanoscale Systems (CNS), a member of the National Nanotechnology Coordinated Infrastructure Network (NNCI), which is supported by the National Science Foundation under NSF award 1541959. K.W. and T.T. acknowledge support from the Elemental Strategy Initiative conducted by the MEXT, Japan and the CREST (JPMJCR15F3), JST. This project has received funding from the European Union's Horizon 2020 research and innovation programme under the Marie Skłodowska-Curie grant agreement No 101023276.

**Author contributions**
E.L.P., T.I.A., G.S., V.W. P.K., H.P. and M.D.L. conceived the project. E.L.P., T.I.A. and G.S. designed and performed the experiments, analyzed the data and wrote the manuscript with extensive input from the other authors. Device fabrication was done by A.Y.J. and A.M.M.V. B.K. and J.H. grew the TMD crystals. T.T. and K.W. grew the hBN crystals. P.K., H.P., and M.D.L. supervised the project.

**Competing interests**
The authors declare no competing interests.

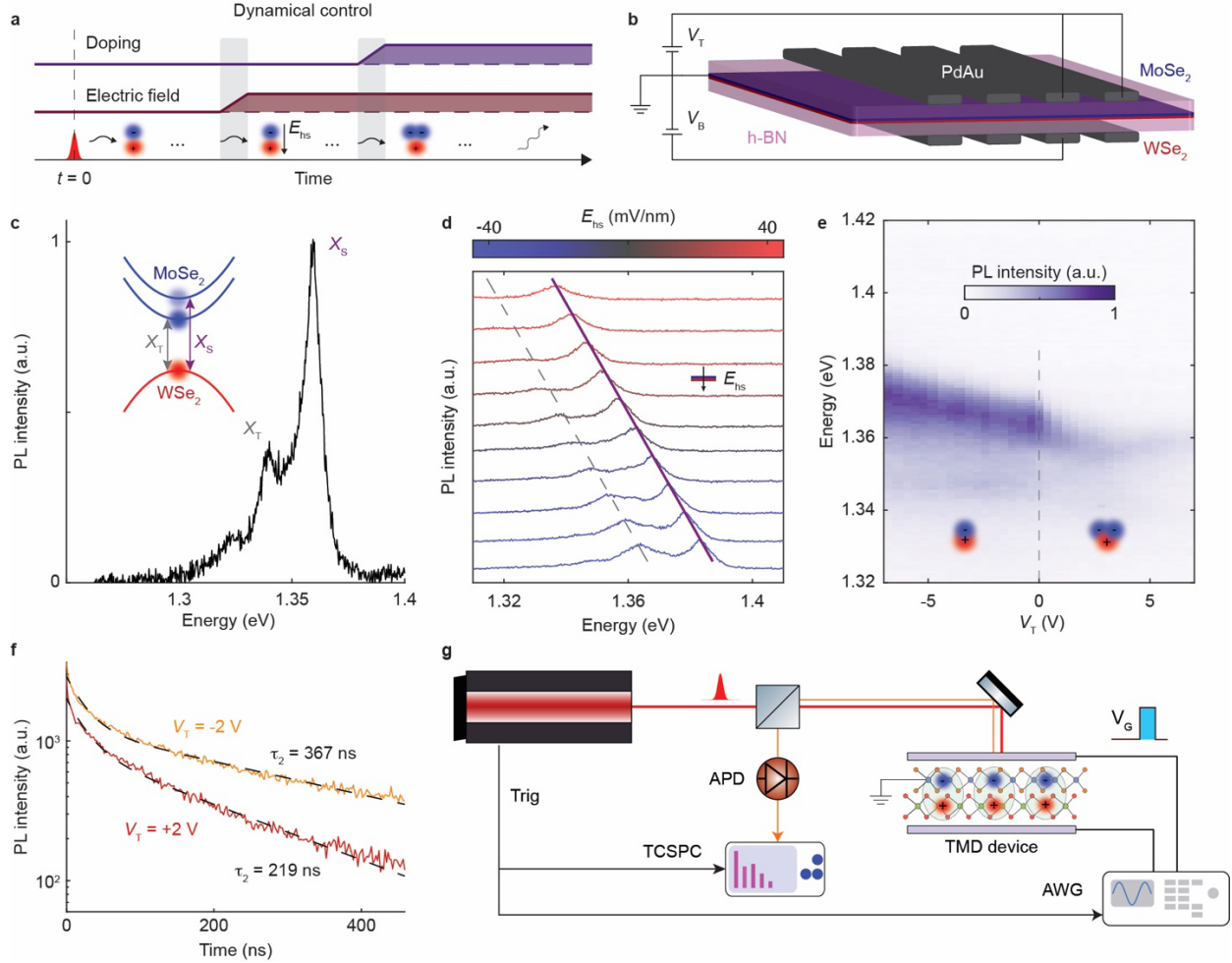

**Fig. 1: Device schematic and gate tunability. a**, Illustration of dynamical exciton control. Electric fields and doping are modulated to tune the properties of excitons between the time of their creation and recombination. **b**, Schematic of MoSe$_2$/WSe$_2$ heterostructure device with patterned gates. The gates are 1 μm wide and laterally separated by about 300 nm. **c**, Low-temperature ($T \approx 6$ K) PL spectrum with both gates grounded. The two most prominent peaks are attributed to the singlet and triplet exciton states. Inset: Band diagram of the MoSe$_2$/WSe$_2$ heterostructure including the upper and lower MoSe$_2$ conduction bands, which correspond to the singlet ($X_S$) and triplet ($X_T$) emission features. **d**, IE PL spectra as a function of vertical electric field. The purple line is a linear fit of the singlet exciton peak energy versus electric field $E_{hs}$. The dashed line through the $X_T$ peaks is a guide to the eye. **e**, Single-gate dependence of the interlayer exciton PL spectrum showing the formation of charged IEs for positive gate voltages. **f**, PL lifetime measurements of the neutral (yellow) and negatively-charged (red) interlayer excitons. Each measurement is averaged for the same amount of time. The slow timescales are indicated in the figure; the fast timescale is about 26 ns in both cases. The sample was excited using a pulsed laser with $\lambda \approx 660$ nm for all measurements, with optical powers of several μW for the spectral measurements and 500 nW for the lifetime measurement in **e**. **g**, Experimental setup for dynamical exciton control. Excitons are generated by a short (~10 ps) laser pulse and PL is collected with an avalanche photodiode (APD) and time-binned with a time-correlated single photon counter (TCSPC).

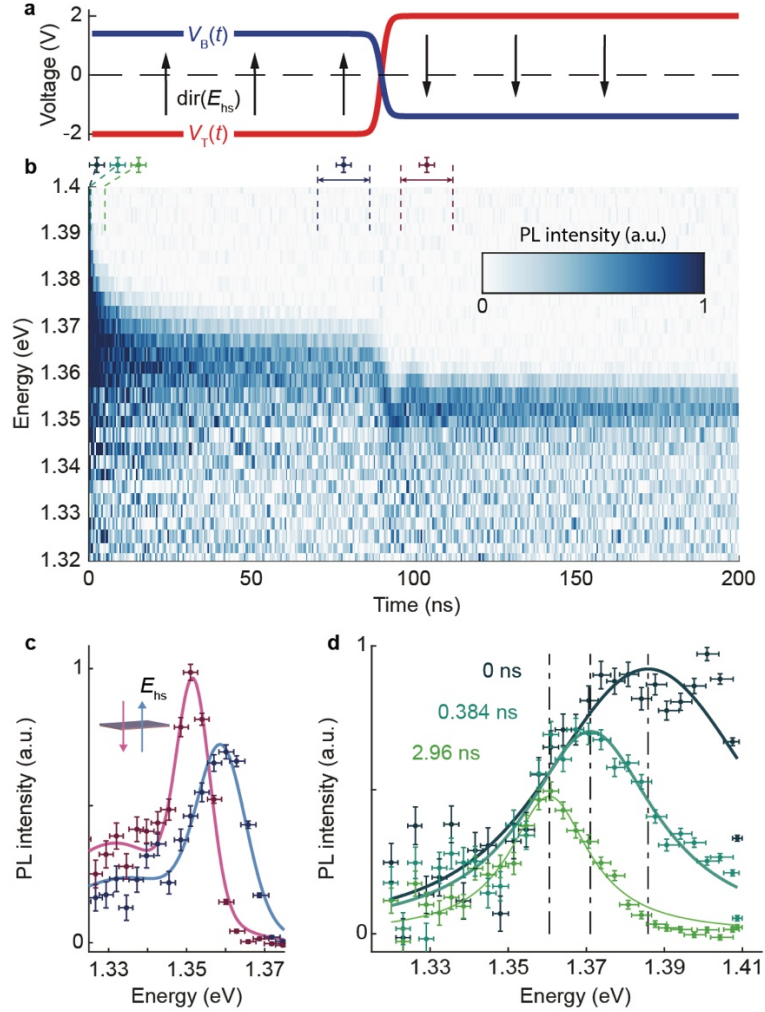

**Fig. 2: Dynamical exciton modulation. a**, Gate sequence for the dynamical tuning of exciton PL energy. The gate voltages are chosen to apply an out-of-plane electric field $E_{hs}$ to the sample, which is reversed within the exciton lifetime. Arrows indicate the direction of the vertical electric field before and after the gate voltages are reversed. **b**, Time-resolved exciton spectrum subject to the gate sequence in **a** and optical excitation power of $P = 500$ nW. Marks above the data indicate time ranges for the linecuts in **c** and **d**. **c**, Photoluminescence spectra integrated over the 16 ns time ranges before (blue) and after (pink) the electric field indicated by crosshairs above **b**. The data are fit to bi-Voigt functions. The vertical errorbars indicate Poissonian shot noise, while horizontal errorbars follow from the finite cutoff width of the tunable filter swept to produce this dataset. Negative intensity values are due to background subtraction (See Supplemental Information). **d**, Relaxation of dipolar blueshift shortly after pulsed excitation at higher power ($P = 1.4$ μW), fitted to Voigt profiles. The spectra here are binned with a time interval $\Delta t = 384$ ps; the beginnings of the time bins are also indicated by the crosshairs above **b**.

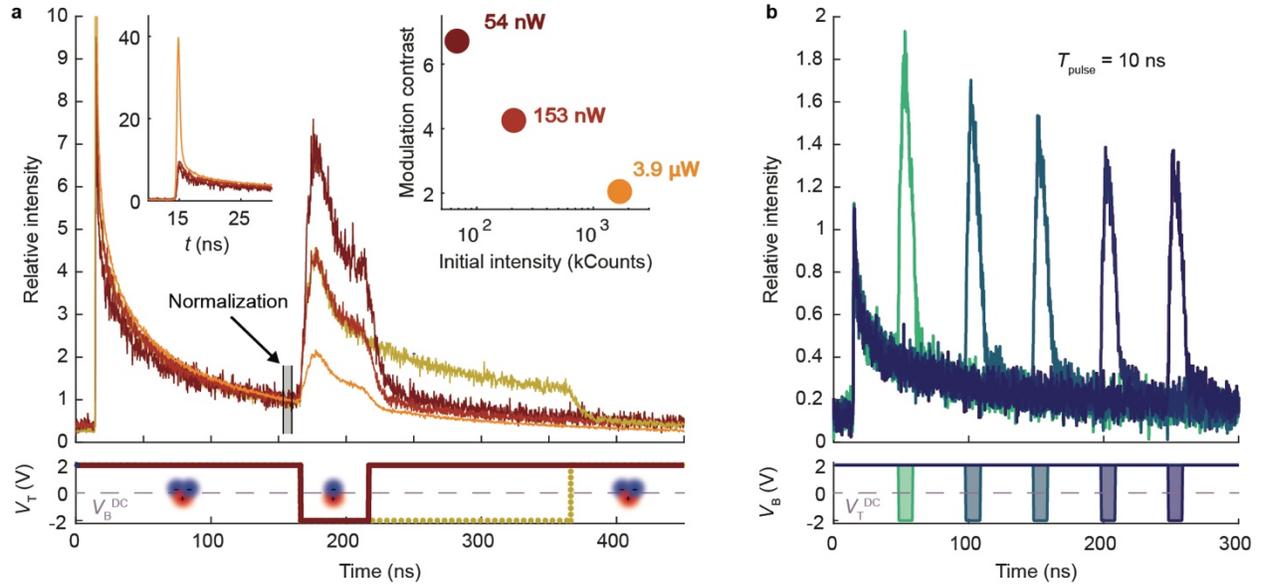

**Fig. 3: Exciton response to pulsed single-gate modulation**. **a**, Top: Exciton PL response to pulsed top gate voltage at varying optical excitation powers with the bottom gate grounded, $V_B^{DC} = 0$. $V_T$ is +2 V at the time of excitation and is toggled to -2 V for 50 ns. To highlight the PL contrast, the data are normalized to the average PL intensity over a short time interval immediately before the gate pulse. The gold curve is taken under identical conditions to the measurement with $P = 153$ nW, except that the gate is held at -2 V for 200 ns instead of 50 ns. Bottom: Pulse sequence. Left inset: initial peak heights showing interaction-driven diffusion at higher excitation power. Right inset: modulation contrast (ratio of PL intensity slightly after/before the gate voltage is toggled) as a function of $P$. **b**, Exciton PL response to 10 ns pulsed bottom gate voltage at very low optical power ($P = 10$ nW) with varying pulse delay times and the top gate grounded, $V_T^{DC} = 0$. The data are normalized to the intensity of the initial optical excitation peak.

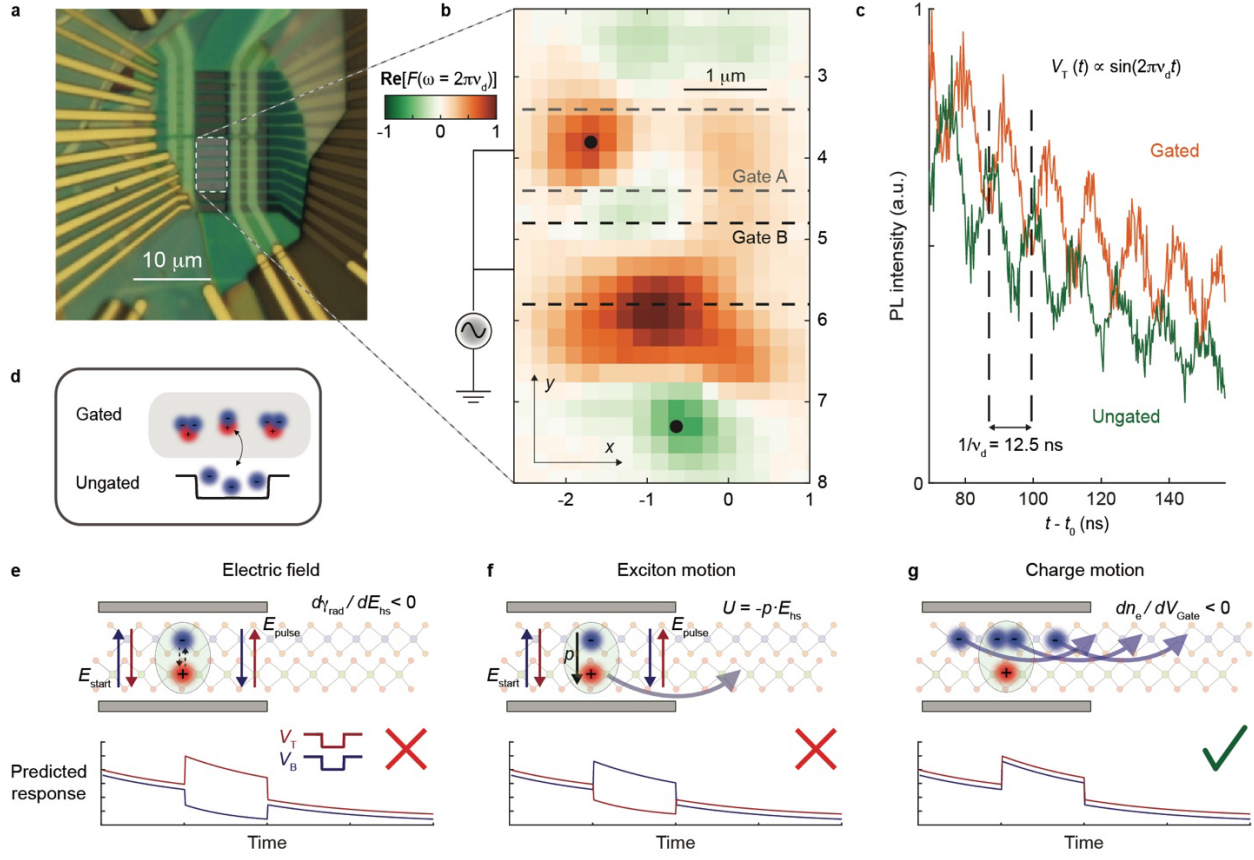

**Fig. 4: Single-gate exciton modulation via charge redistribution. a**, Device micrograph. The device is patterned with 13 top and bottom gates, which are laterally aligned to each other, and of which 2 are operated in this measurement. **b**, Spatial map of the interlayer exciton response to an a.c. input gate voltage applied to two adjacent top gates, $V_{T, (Gate\ A)}(t) = V_{T, (Gate\ B)}(t) = (+2\ V) \sin 2\pi \nu_d t$, where $\nu_d = 80$ MHz. The real part of the Fourier transform at the drive frequency, $\text{Re}[F(x, y, \omega = 2\pi\nu_d)]$, has opposite signs underneath and outside of the gates. The imaginary part is small, indicating that the complex phase of the response is predominantly 0° or 180° where the response is strong. **c**, Raw data at two representative sample locations indicated by black dots in **b** illustrating out-of-phase responses. **d**, Schematic illustration of proposed exciton modulation response origin. Neighboring charge traps around the connected gates harbor reservoirs of free electrons, which are shuttled underneath the gate by the fast modulation to produce charged interlayer excitons. **e,f,g** Predicted exciton response from different hypothetical control mechanisms, e.g. electric field control of radiative rate (**e**), trapping of excitons in a gate-defined potential (**f**), and change of exciton character due to charge control (**g**). Here, $\gamma_{rad}$ refers to the radiative decay rate, $U$ refers to the Stark potential of interlayer excitons in an out-of-plane electric field, and $n_e$ refers to the electron density in the MoSe$_2$ layers. Only the final mechanism, in which electrons are moved out of the gated region when $V_T$ or $V_B$ becomes negative (and vice-versa), is consistent with our observations.

# Supplementary Materials for

# Dynamical Control of Excitons in Atomically Thin Semiconductors

Eric L. Peterson[1*], Trond I. Andersen[1,2*], Giovanni Scuri[1,3], Andrew Y. Joe[1,4], Andrés M. Mier Valdivia[5], Xiaoling Liu[1], Alexander A. Zibrov[1], Bumho Kim[6], Takashi Taniguchi[7], Kenji Watanabe[8], James Hone[9], Valentin Walther[1,10,11], Hongkun Park[1,12], Philip Kim[1,5], Mikhail D. Lukin[1,†]

[1]Department of Physics, Harvard University, Cambridge, MA 02138, USA
[2]Google Research, Mountain View, CA 94043, USA
[3]E. L. Ginzton Laboratory, Stanford University, Stanford, CA 94305, USA
[4]Department of Physics and Astronomy, University of California, Riverside, CA 92521, USA
[5]John A. Paulson School of Engineering and Applied Sciences, Harvard University, Cambridge, MA 02138, USA
[6]Department of Physics and Astronomy, University of Pennsylvania, Philadelphia, PA 19104, USA
[7]International Center for Materials Nanoarchitectonics, National Institute for Materials Science, 1-1 Namiki, Tsukuba 305-0044, Japan
[8]Research Center for Functional Materials, National Institute for Materials Science, 1-1 Namiki, Tsukuba 305-0044, Japan
[9]Department of Mechanical Engineering, Columbia University, New York, NY 10027, USA
[10]Department of Chemistry, Purdue University, West Lafayette, IN 47906, US
[11]Department of Physics and Astronomy, Purdue University, West Lafayette, IN 47906, USA
[12]Department of Chemistry and Chemical Biology, Harvard University, Cambridge, MA 02138, USA

*These authors contributed equally to this work.
†To whom correspondence should be addressed: lukin@physics.harvard.edu

**Contents:**

Supplementary Notes I-VIII
Supplementary Figs. S1 to S6

**Supplementary Note I.    PL Relaxation Dynamics**

The interlayer exciton relaxation dynamics in Device A (Fig. 1f) are clearly not described by a simple exponential decay. Such non-exponential relaxation behavior has been routinely (though not universally) observed in photoluminescence from MoSe$_2$/WSe$_2$ heterobilayers. Various studies have observed and fit the exciton photoluminescence decay curve to a single exponential[1–4], a double exponential[5–8], or a triple exponential[9,10] function, with explanations of the underlying dynamics ranging from singlet and triplet excitons, neutral and charged excitons, mixing between direct and indirect excitons, to exciton-exciton interactions and sample inhomogeneity. Many of these factors can in principle be sensitive to strain and small variations in twist angle, hence sample-to-sample variation makes unambiguous assignment of the different timescales challenging.

Complex relaxation processes in solids have also been described using a stretched exponential function, i.e. a function of the form

$$I(t) = A\exp[-(t/\tau_{SE})^\beta], \quad (S1)$$

where $0 < \beta \leq 1$[11]. Stretched exponential relaxation in solid-state photoluminescence experiments has been explained in terms of trapping-detrapping processes between localized and delocalized states, or as a continuous sum of simple exponential decays[12]. Outside of a lone study of diffusion in WS$_2$/tetracene heterostructures[13], the possibility of understanding the dynamics of long-lived excitons in TMDs in terms of such processes has not attracted significant attention, though it could conceivably arise from mid-gap states or diffusion through moiré potentials. In Fig. S1a, we compare the stretched exponential to the biexponential function that has typically been used to fit interlayer exciton PL lifetime data, and find that the stretched exponential function better captures the form of the decay curve despite having fewer free parameters than the biexponential function.

As in the other time-resolved measurements presented in this work, the measurement in Figure S1 is carried out using a pulsed laser with a repetition period of $T \approx 474$ ns. To account for any residual exciton population at the onset of a new pulse, we include contributions from past pulses by fitting the data to a sum,

$$Y(t) = \sum_{n=0}^{9} y(t + nT), \quad (S2)$$

where $y(t)$ is either the biexponential or stretched exponential function to be fit.

Following previous works[12,14,15], the stretched exponential function can be expressed as a continuous sum of simple exponential functions,

$$\exp[-(t/\tau_{SE})^\beta] = \int_0^\infty P_\beta(s) e^{-s\Gamma^* t} ds \quad (S3)$$

where $s \equiv \Gamma/\Gamma^*$, $\Gamma^* \equiv 1/\tau_{\text{SE}}$, and $\Gamma$ is a specific relaxation rate. Noting that $P_\beta(s)$ is therefore given by the inverse Laplace transform of $\exp[-(t/\tau_{\text{SE}})^\beta]$, the distribution of relaxation rates can ultimately be expressed as

$$P_\beta(s) = \frac{1}{\pi}\int_0^\infty e^{-u^\beta \cos(\pi\beta/2)} \cos[su - u^\beta \sin(\pi\beta/2)]\, du \tag{S4}$$

which we evaluated numerically for comparison with the biexponential model (Fig. S1b).

We also note that in behavior deviating from simple exponential decay, the exciton *population* decay is not described by the same functional form as the measured *photoluminescence* decay. In general, the PL intensity is proportional to the exciton population decay rate, $I(t) = -\eta \dot{n}(t)$, where $n$ is the exciton population and $\eta$ is a proportionality factor accounting for the device's quantum efficiency and the collection efficiency of the optical setup (the minus sign accounts for the fact that $n$ is strictly decreasing, whereas $I$ is strictly positive). Assuming that $\eta$ is constant in time, integration gives

$$\frac{n(t)}{n_0} = 1 - \frac{1}{\eta n_0}\int_0^t I(t')dt' \tag{S5}$$

Carrying out the integral for the case of biexponential decay with $I(t) = A_1 e^{-t/\tau_1} + A_2 e^{-t/\tau_2}$, and applying the constraint that $n(t \to \infty) = 0$, the population decay is given by

$$n_{\text{Bi}}(t) = \frac{1}{\eta}\left(A_1\tau_1 e^{-t/\tau_1} + A_2\tau_2 e^{-t/\tau_2}\right) \tag{S6}$$

where $A_1\tau_1 + A_2\tau_2 = \eta n_0$, which follows from the $n(t \to \infty) = 0$ constraint. Eqn. (S6) implies that the proportion of excitons which are long-lived (i.e. associated with the long timescale $\tau_2$) is much more favorable than the raw PL decay data suggests. Specifically, the two decay components are now weighted by $A_1\tau_1$ (fast decay) and $A_2\tau_2$ (slow decay) rather than just $A_1$ and $A_2$. In our case, since $\tau_2 \gg \tau_1$, accounting for the reweighted coefficients in Eqn. (S6) leads to the conclusion that a significant majority of the exciton population in our sample decays over the slow timescale (Fig. S1b) rather than the fast one in the biexponential model.

In the case of the stretched exponential model, integrating the PL decay curve in Eqn. (S4) transforms the decay rate distribution for the exciton population from $P_\beta(s)$ to $s^{-1}P_\beta(s)$, which again skews the distribution towards slower decay rates (Fig. S1b).

The main consequence of the non-exponential decay for this study is that it becomes more challenging to fully understand the gate-driven exciton dynamics when the exciton decay rate is not constant in the DC case. Developing a more thorough and rigorous understanding of exciton dynamics in twisted TMD bilayers is an ongoing research effort that will benefit from additional systematic study and probes which are complementary to those used in this work[16,17].

**Supplementary Note II.    Time-Resolved Spectrum Reconstruction**
Here we describe the time-resolved spectroscopy technique used to collect and analyze the data in Figure 2. We perform time-resolved spectroscopy by sweeping an angle-tunable shortpass filter through the interlayer exciton spectrum and collecting time-resolved PL data at each filter angle.

To calibrate the filter angles $\theta_i$ to wavelengths $\lambda_i$, the angle is first swept through a known reference spectrum $S(\lambda)$ from a halogen lamp to produce a collection of filtered spectra $S_{\theta_i}(\lambda)$ representing the reference spectrum through an array of filter cut-off wavelengths given by the filter angles. The spectral transmission that is uniquely captured by each angle $\theta_i$ is then

$$s_i(\lambda) = S_{\theta_i}(\lambda) - S_{\theta_{i+1}}(\lambda), \tag{S7}$$

From the definition of $s_i(\lambda)$, it is clear that calibrating $N$ wavelengths requires measuring the reference spectrum at $N+1$ filter angles. We take $\lambda_i$ to be the maximum of $s_i(\lambda)$ and use the FWHM of $s_i(\lambda)$ as horizontal errorbars in representing the spectral data after processing. Finally, we apply a correction factor $\alpha_i = s_i(\lambda_i)/S(\lambda_i)$ to account for filter angle-dependent spectral distortion in the subsequent analysis. These steps are illustrated in Fig. S2a.

After calibrating the tunable shortpass filter angles, we repeat a time-resolved exciton PL measurement at each filter angle (Fig. S2b), collecting $N+1$ PL measurements $I_{\theta_i}(t)$. We reconstruct the full time-resolved spectrum by isolating the photon counts corresponding to each angle $\theta_i$ as in Eqn. (S7), assigning them to the calibrated wavelength $\lambda_i$, and applying the correction factor $\alpha_i$, resulting in the time-resolved spectrum

$$I(\lambda_i, t) = \alpha_i \cdot \left[I_{\theta_i}(t) - I_{\theta_{i+1}}(t)\right]. \tag{S8}$$

The vertical errorbars in Figs. 2c-d represent Poissonian shot noise. We add the noise from the two PL measurements that are subtracted to reconstruct each wavelength, so that the errorbars are given by

$$\sigma(\lambda_i, t) = \sigma_{\theta_i}(t) + \sigma_{\theta_{i+1}}(t) = \sqrt{\alpha_i} \cdot \left(\sqrt{I_{\theta_i}(t)} + \sqrt{I_{\theta_{i+1}}(t)}\right), \tag{S9}$$

**Supplementary Note III.    Characterization of high-speed electrical transmission**
In our gate-modulated spectral measurements, we observed a damped oscillation of the exciton PL energy after the gate voltages are changed suddenly (Fig. 2b). Here we explain how this is related to the high-frequency transmission characteristics of our device and cryogenic measurement setup.

The gate voltages used to produce the electric field reversal in Fig. 2b are described by a step function, $V_{T,B}(t) = V_{T,B}^0 \left(1 - 2\theta(t - t_R)\right)$, where $t_R$ is the time of the electric field reversal and $V_{T,B}^0$ are their voltages at the start of the measurement. When subjected to a step-function input, underdamped oscillators (which describe RLC circuits for some parameter combinations) generically respond with an exponentially damped oscillation (known as a "ringdown" oscillation) at their resonance frequency. The oscillation observed in Fig. 2b is suggestive of this behavior, with an apparent resonance frequency of about 80 MHz.

In order to understand this behavior, we characterized the RF transmission through our cryostat feedthroughs and into the device by measuring the response of the exciton PL intensity at various frequencies, several of which are plotted in Figure S3. Although this method is not as direct as using a network analyzer because the transmitted signal is convolved with the voltage dependence of the exciton PL intensity (which is not strictly linear), it is more thorough as the transmission through the full device is manifested in the exciton response. The data show that the top gate and bottom gate have transmission maxima at about 80 MHz and 100 MHz respectively (Figs. S3a-b), with the top gate resonance being narrower in frequency (characteristic of underdamped resonators). The 80 MHz resonance associated with the top gate coincides with the ringdown frequency observed in Fig. 2b; we plot several of time-resolved PL measurements used to construct the spectra in Fig. 2b for comparison (Fig. S3c).

The RF transmission characteristics appear to be due to factors external to the van der Waals heterostructure itself. If they were determined primarily by the device itself, then the timescales involved would depend on the resistances of the TMD layers in the doping modulation experiments (Fig. S3a) and of the top and bottom gates in the electric field modulation experiments (Fig. S3c). These are expected to be very different; however, the doping and electric field experiments both show a similar resonance around 80 MHz. As further evidence that the RF transmission at high frequencies is limited by factors extrinsic to the TMD device, we observe transmission properties in a second device (Device B) which are similar to those of the bottom gate of the device studied in the main text (Device A) despite their very different geometries (Fig. S3d).

Knowledge of the top and bottom gate RF transmission behavior informs our understanding of the electrical ringdown evident in our time-resolved data. The top and bottom gates have different RF transmission properties, and the electric field that produces the oscillation in PL energy is proportional to both top and bottom gate voltages, so the RF transmission to each gate has to be accounted for in combination. The top gate transmission resonance frequency matches the ringdown oscillation frequency (Fig. S3a, c), while the bottom gate transmission maximum is of a different frequency, so we attribute the ringdown oscillation to the top gate alone. The transient response of the bottom gate is nevertheless evident in the mean PL energy as a function of time after the electric field reversal (Fig. S4) – in particular, if the top gate alone were responsible for the transient, then the ringdown oscillation amplitude would be equal to the change in equilibrium PL energy before and after the field reversal, but the oscillation amplitude is clearly smaller than the total change in energy. A full model of the PL energy ringdown, including the transient responses of both gates, is given by

$$E(t) = E_i \theta(t + t_R) + \theta(t - t_R)[E_f + A e^{-\sigma_{\text{TG}}(t-t_R)} \cos(2\pi f_{\text{TG}}(t - t_R) + \varphi) \\ + (E_f - E_i - A \cos \varphi) e^{-\sigma_{\text{BG}}(t-t_R)}] \qquad (S10)$$

where $\theta(t - t_R)$ is the Heaviside step function, $E_i$ and $E_f$ are the PL energies before and after the voltage step pulse, $\sigma_{\text{TG}}$ and $\sigma_{\text{BG}}$ are the decay times associated with the top and bottom gate transient responses, and $A$ and $\varphi$ are the amplitude and phase of the ringdown oscillation. The prefactor in the bottom gate transient term follows from continuity of $E(t)$ at $t = t_R$. Fitting the mean PL energy in Fig. 2b to this model (Fig. S4, blue curve) we extract $\sigma_{BG}^{-1} \approx 8.6$ ns, $\sigma_{TG}^{-1} \approx 14.8$ ns, and $f_{TG}^{-1} \approx 12.4$ ns. The combined gate operation time in our experiments is therefore limited by the top gate transient response time of about 15 ns.

Fitting the same data to a simpler model that neglects the bottom gate transient (Fig. S4, red curve) yields similar results: $\sigma_{TG}^{-1} \approx 15.3$ ns, and $f_{TG}^{-1} \approx 11.7$ ns. This fit model is given by

$$E(t > t_R) = E_f + Ae^{-\sigma_{TG}(t-t_R)} \cos(2\pi f_{TG}(t - t_R) + \varphi) \tag{S11}$$

**Supplementary Note IV.    Estimate of Device RC Constant**

Although the gate response time seems to be limited by factors external to the device, it is still instructive to estimate a device RC constant based on a toy model. Expressing the resistance and geometric capacitance of our two-dimensional sample as

$$R = R_{sh} \frac{L}{2W}, \tag{S12}$$

$$C = \varepsilon_{hBN} \frac{LW}{d} \tag{S13}$$

where $L \approx 1$ µm is the gate width (which is also approximately the distance for charge to travel between gated an ungated regions), $W$ is the gate length (which is also related to the areal cross-section for charge to flow into the gated region), $d$ is the hBN thickness (100 nm for the top gate, and 70 nm for the bottom gate), and $R_{sh}$ is the sheet resistance of the TMD heterobilayer. Multiplying these expressions gives a device RC constant of

$$RC = \varepsilon_{hBN} R_{sh} \frac{L^2}{2d} \tag{S14}$$

In this estimation, $RC$ does not depend on the gate length $W$. This is intuitively consistent with our proposed picture in which charge is pulled in from either side of the gate, a process which in our gate geometry should depend on the width (but not the length) of the gate.

From Eqn. (S14), we can estimate an upper bound on the sheet resistance necessary to drive the device with a given time constant. In particular, we find that for a device with our back gate geometry, a time constant of $RC = 10$ ns corresponds to a sheet resistance $R_{sh} \approx 41$ MΩ/sq.

**Supplementary Note V.    Second device with similar measurements**

We have additionally demonstrated dynamical modulation of interlayer excitons in a second MoSe$_2$/WSe$_2$ device (Device B) featuring a single monolithic gate, in contrast to the patterned array of gates in the device described in the main text (Device A). The gate-dependent spectra (Fig. S5a) and PL modulation response (Fig. S5b) of Device B are discussed here.

We modulated the interlayer excitons in this device by applying a DC-biased RF voltage to the gate, $V_G(t) = V_{DC} + (2\text{ V})\sin[(104\text{ MHz})2\pi t]$, and collecting the emitted PL for several values of $V_{DC}$. The data in Fig. S5 show that dynamical exciton control is also possible in this device, although the amplitude of the response is much lower than in Device A. The modulated PL data in Device B is well-fit by a sinusoidally-modulated exponential decay curve,

$$I(t) = Ae^{-t/\tau}(1 + B\sin(2\pi f_0 t + \phi_0)), \tag{S15}$$

where $f_0 = 104$ MHz, $\phi_0$ is a constant phase that is fixed between all measurements in Fig. S5b and $A$, $B$, and $\tau$ are free parameters.

The data confirms that high-speed exciton manipulation is reproducible in multiple devices. Although the quantum efficiency in this device was not high enough to measure the response at very low optical excitation powers, the lower amplitude of the PL modulation is consistent with the response being electric field-driven rather than charge-driven in this device, as expected for a device with a monolithic gate.

**Supplementary Note VI.    Effect of Gate Voltage in Ungated Regions**
A key observation in this work was that neighboring regions near the gate edge have opposite responses to fast gate modulation, and that ungated regions do not respond to quasistatic changes in gate voltage (Fig. 4). Here we elaborate on this.

Figure S6 shows the contrasting DC and RF responses at such a pair of spots. In Fig. S6a, a 50 ns pulse toggles the top gate from +2 V to -2 V for 50 ns and a time-resolved PL measurement is performed at each sample location. In Figs. S6b-c the top gate is slowly swept from -7 to +7 V at a rate of 0.1 V every 4 minutes, and a spectrum at each sample location is acquired at each voltage The DC and RF measurements show qualitatively different behavior: the PL spectrum of the gated spot (Fig. S6b) shows a slight Stark shift for negative voltages and a transition from the neutral to charged exciton states around 0 V, which manifests as a redshift and weakened intensity for positive voltages; by contrast, the ungated spot (Fig. S6c) shows no discernable response to the slow gate sweep, but an RF response which is similar in magnitude to the gated spot.

Our data suggests that the device charging behavior depends nontrivially on the rate at which the gate voltages are modulated: when the voltage changes slowly, the $MoSe_2$ remains in electrochemical equilibrium with the ground that it is connected to, and charge is pulled into the gated region from the metallic contact to the TMD flake without affecting the ungated regions. When the voltage changes more quickly than charge can be injected from the contact, the $MoSe_2$ is no longer in electrochemical equilibrium with the contact and charge is pulled from neighboring regions in order for the gated region to respond to the change in gate voltage.

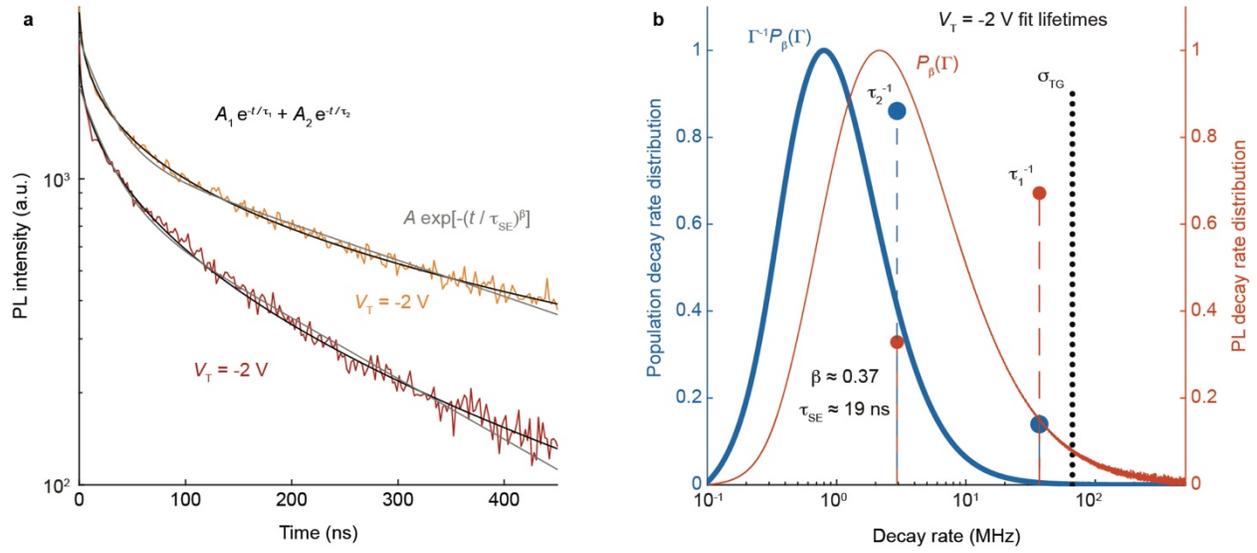

**Fig. S1: Biexponential and stretched exponential exciton decay models**. **a**, Biexponential (grey) and stretched exponential (black) fits to the intrinsic and n-doped PL lifetime measurements shown in Fig. 1f. The stretched exponential function fits the data slightly better, with one fewer fit parameter. **b**, PL (orange) and population (blue) decay rate distributions in the biexponential model (vertical dashed lines) and stretched exponential model (continuous curves) for the intrinsic $V_T = -2$ V dataset in **a**. The amplitudes of the biexponential decay components are normalized to sum to 1, while the stretched exponential rate distributions are themselves normalized by their peak value. The stretched exponential distributions $P_\beta(s)$ are expressed in terms of the physical decay rate $\Gamma = s\Gamma^* = s/\tau_{SE}$. The top gate response time $\tau_{TG}^{-1} = \sigma_{TG}$ extracted from Fig. S4 is also indicated by a dashed vertical line for comparison to the intrinsic exciton dynamics in the device.

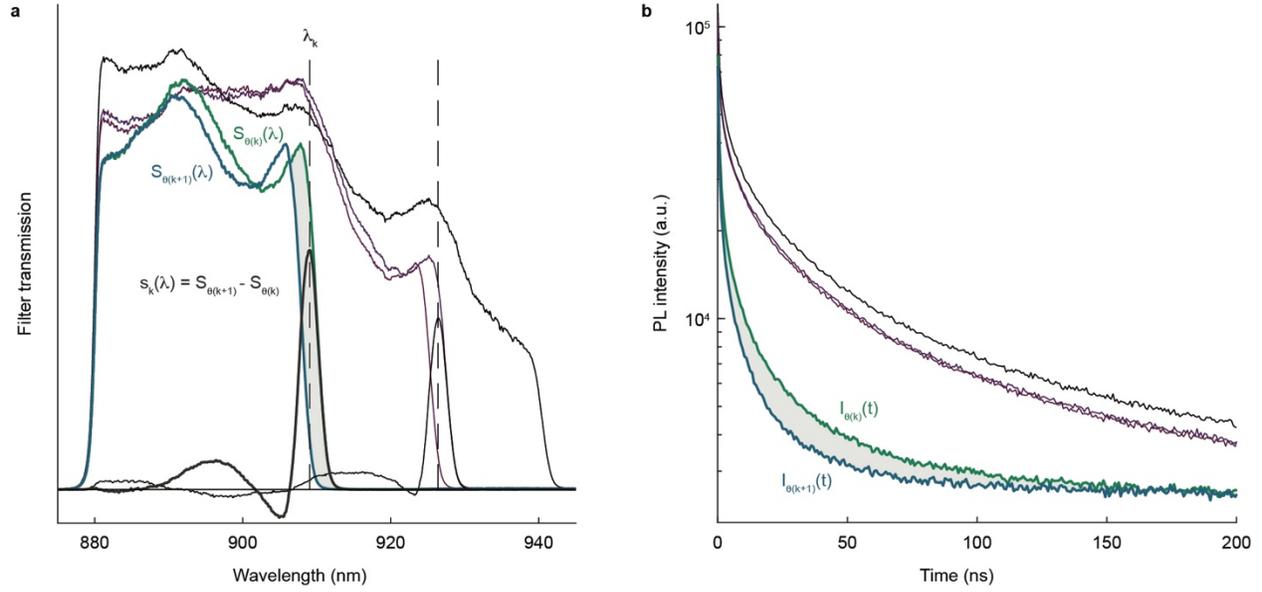

**Fig. S2: Illustration of Time-Resolved Spectroscopy Method. a**, Calibration of shortpass filter to reference halogen lamp spectrum. As an example, we show the reference spectrum at five different filter angles: one with the cutoff at ~940 nm (far away from the IE resonance), and two pairs of neighboring filter angles at intermediate cutoff wavelengths. The spectral windows $s_i(\lambda)$ computed according to Eq. (S7) are plotted on top of the neighboring filtered reference spectra. The spectral windows $s_i(\lambda)$ are not perfectly symmetric or strictly positive because the filter slightly modifies the transmission away from the cutoff wavelength. **b**, Example time-resolved IE photoluminescence measurements at the same filter angles as in **a**. The area between the $\theta_k$ and $\theta_{k+1}$ curves are shaded to indicate the contribution to the signal at the wavelength about the similarly-shaded region in **a**. The much smaller separation between the other pair of neighboring curves reflects that the IE PL is much weaker in that spectral window.

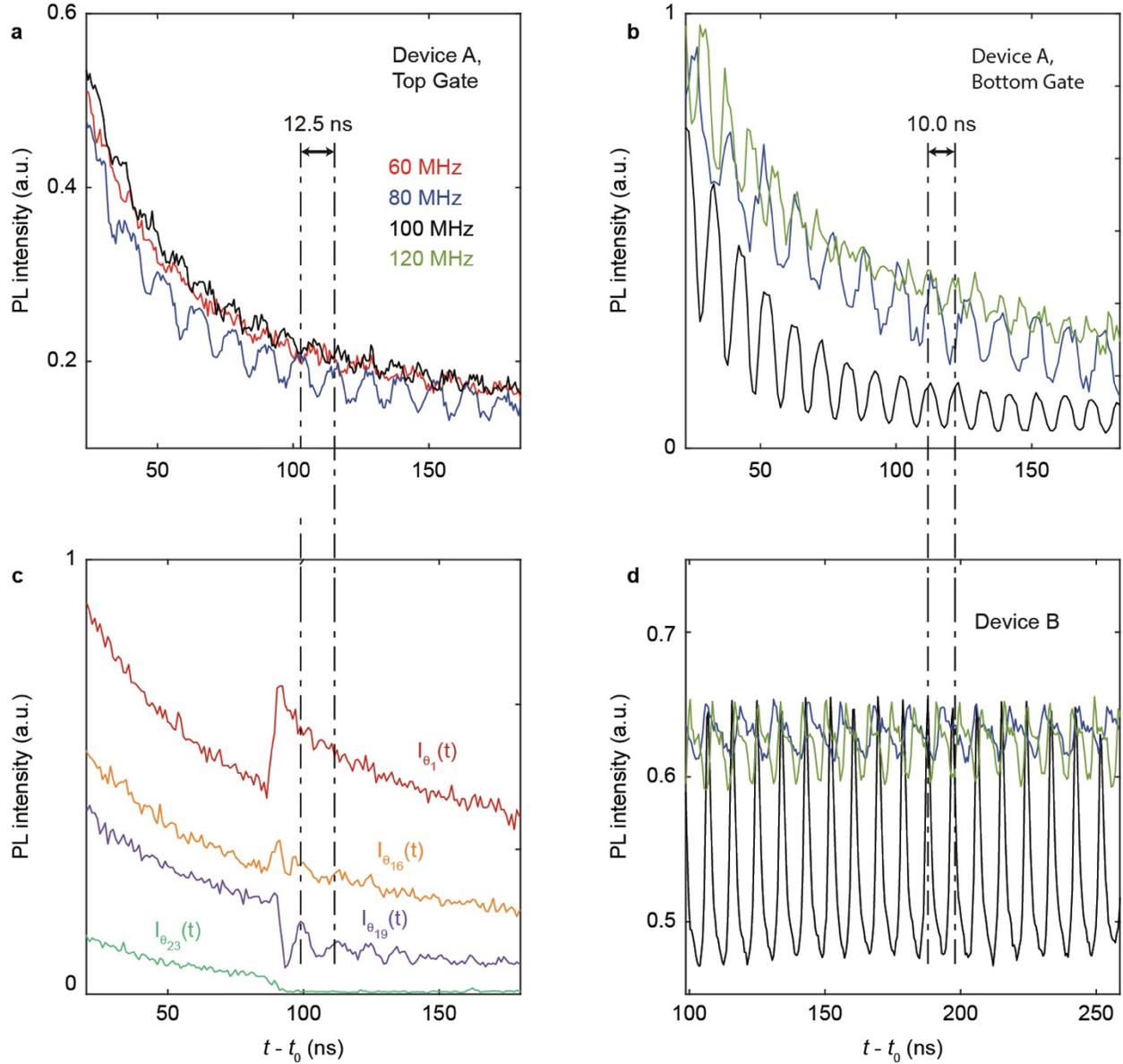

**Fig. S3: Radiofrequency transmission to top and bottom heterostructure gates**. **a**, Top gate-modulated exciton PL measurements in Device A (the device studied in the main text). The response to gate modulation is strong at 80 MHz but relatively weak at 60 MHz and 100 MHz, suggesting resonant RF transmission. **b**, Bottom gate-modulated exciton lifetime measurements in Device A. The response is strongest at about 100 MHz, in contrast to the 80 MHz resonance of the top gate. Additionally, the response is appreciable tens of MHz away from the resonance, unlike the relatively sharp RF transmission maximum of the top gate. **c**, Selected raw time-resolved PL measurements used to construct the time-resolved PL spectra in Fig. 2b. **d**, *Intralayer* exciton response at 80 MHz, 100 MHz, and 120 MHz in Device B, the device studied in Fig. S5. Again, there is a resonance at around 100 MHz.

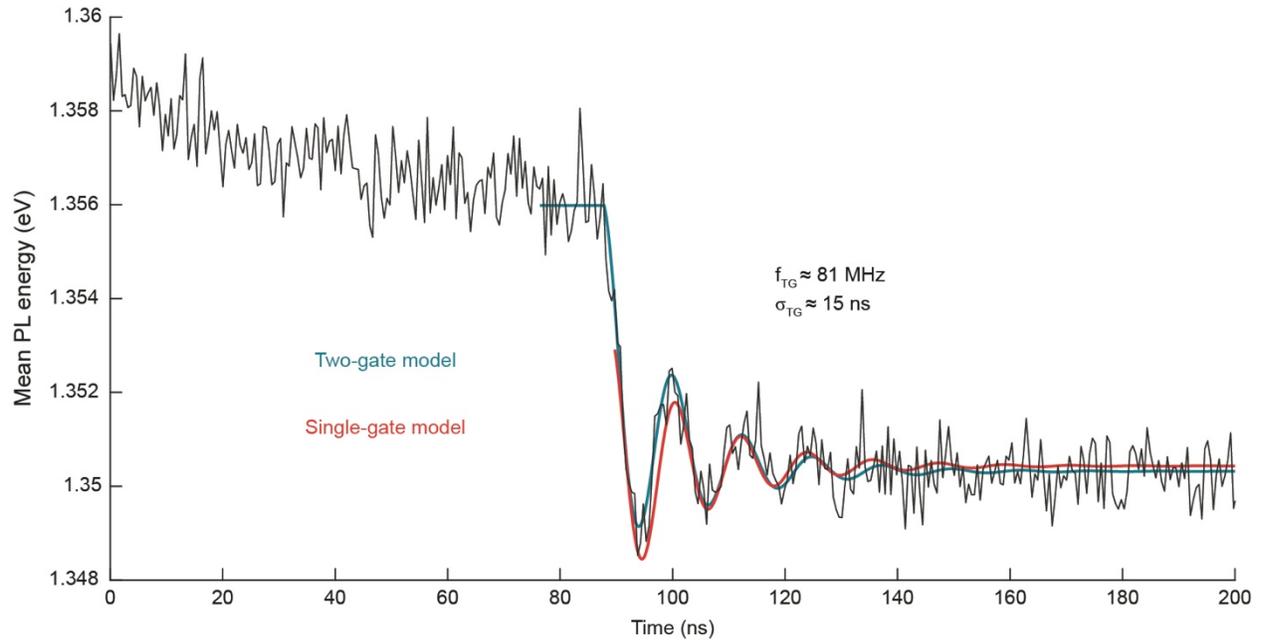

**Fig. S4: Transient gate response**. Black curve: mean time-resolved PL energy from the dynamical PL energy modulation data in Fig. 2b. Blue curve: fit to two-gate transient model in Eq. (S10). Red: fit to single-gate transient model in Eq. (S11). Both models give similar results for the top gate RF transmission resonance and transient decay time, which are inset in the figure.

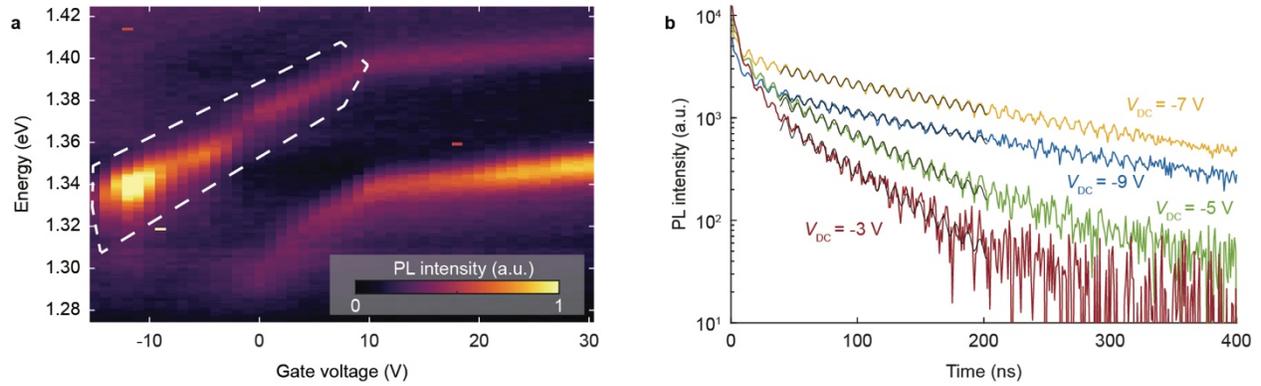

**Fig. S5: Demonstration of control scheme in second device. a**, Gate-dependent interlayer exciton PL spectrum in Device B. The feature enclosed in the white dashed line is attributed to interlayer exciton emission. **b**, PL intensity response to RF gate modulation (104 MHz) at several bias voltages in Device B. The natural PL decay curve in this device is more clearly biexponential in nature. The data after the fast relaxation fits well to a sinusoid with an exponential envelope.

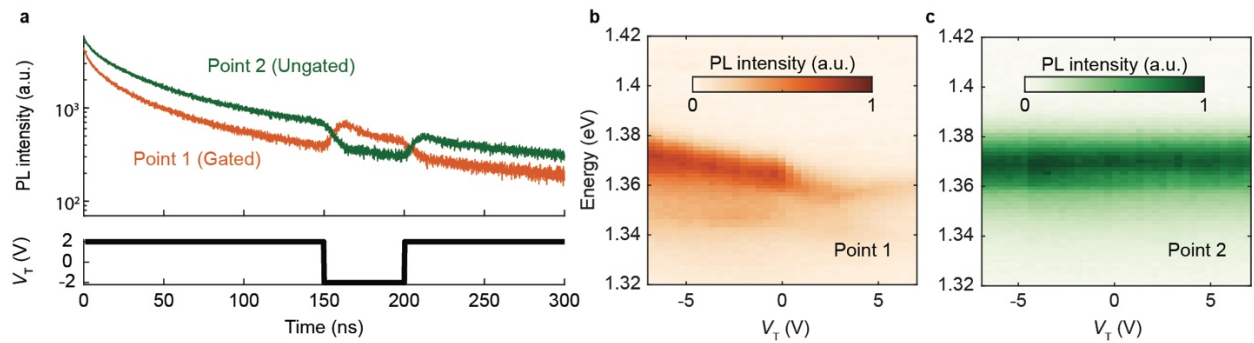

**Fig. S6: Contrasting responses to quasistatic and radiofrequency gate modulation. a**, Dynamical PL intensity response to a 50 ns gate pulse at two neighboring locations on the device, one which is gated and one which is not. The responses are opposite. **b,c** Quasistatic gate voltage dependence of the interlayer exciton PL spectrum at Points 1 and 2 of **a**. In both measurements, spectra were acquired for 4 minutes per voltage point – long enough for charge to be injected through the $MoSe_2$ contact. As expected for a device in electrochemical equilibrium with the ground, the gated region shows a transition from neutral IE to charged IE at $V_T = 0$, while the ungated region is unresponsive to changes in gate voltage.